\documentclass[pre,twocolumn,floatfix,showpacs]{revtex4}
\usepackage{amsmath,amssymb,graphicx,psfrag}

\begin{document}

\title{Phase-field crystal modeling of equilibrium bcc-liquid interfaces}

\author{Kuo-An Wu\footnote{Present address: Department of Materials Science and 
Engineering, Northwestern University, Evanston IL, 60208, USA}} \author{Alain Karma}

\address{
Department of Physics, Northeastern University, Boston, Massachusetts 02115
}

\begin{abstract}
We investigate the equilibrium properties of 
bcc-liquid interfaces modeled with a continuum
phase-field crystal (PFC) approach
[K. R. Elder and M. Grant, Phys. Rev. E \textbf{70}, 051605 (2004)]. 
A multiscale analysis of the PFC model is carried out which exploits the fact 
that the amplitudes of crystal density waves decay slowly into the liquid
in the physically relevant limit where the 
freezing transition is weakly first order. This analysis yields a set of coupled 
equations for these amplitudes that is similar to the set of equations derived 
from Ginzburg-Landau (GL) theory  [K.-A. Wu \emph{et al.}, Phys. Rev. E 
\textbf{73}, 094101 (2006)]. The two sets  only differ in the details of higher 
order nonlinear couplings between different density waves, which is determined by 
the form of the nonlinearity assumed in the PFC model and by the ansatz that all 
polygons with the same number of sides have equal weight in GL theory. Despite 
these differences, for parameters (liquid structure factor and solid density wave 
amplitude) of Fe determined from molecular dynamic (MD) simulations, the PFC and 
GL amplitude equations yield very similar predictions for the overall magnitude 
and anisotropy of the interfacial free-energy and density wave profiles. These 
predictions are compared with MD simulations as well as 
numerical solutions of the PFC model. 
 
\end{abstract}


\maketitle{}

\section{Introduction}
\label{sec:intro}
The phase-field method is by 
now well-developed to simulate the continuum scale evolution of  
interfaces outside of equilibrium with application to
solidification \cite{Boeetal02} and other materials science
problems  \cite{Che02,Kar05}. 
The method rests on a coarse-graining procedure that 
smears out the discrete atomic nature of the interface. Hence, 
the phenomenological form of the 
free-energy functional used to construct a conventional phase-field
model generally needs to be tailored to reproduce quantitatively
atomistically determined interfacial  properties.

An important property in a crystal growth context is 
the anisotropy of the excess free-energy of the
crystal-melt interface that
is a key parameter controlling dendritic evolution 
\cite{Lan87,Kesetal88,BenBre93,KarRap9698,Provetal98,Hoyetal03,Haxetal06}. 
This anisotropy is traditionally incorporated phenomenologically
into the phase-field model by letting the free-energy density
depend on the direction normal to the interface, itself expressed
in terms of the gradient of the phase-field
\cite{ZiaWal85,Macetal93,WheMcF96}.
Computationally efficient implementations of these
models have been successfully applied to simulate
dendritic evolution in materials with both
atomically rough \cite{KarRap9698,Provetal98,Haxetal06} and 
faceted \cite{Debetal03} interfaces. The conventional 
phase-field approach, however, 
falls short in problems where crystalline defects
have a profound influence on morphological evolution. 
For example, solidification twins
can dramatically alter 
both eutectic \cite{DayHel68,Napetal2004} and dendritic \cite{Henetal04} 
microstructures, and crystalline defects ultimately 
control grain coalescence and microstructural evolution 
during and after the late stages of solidification \cite{Rapetal2002}. 

Over the last few years, the phase-field crystal (PFC) method has
emerged as an attractive computational approach to tackle this
class of problems where atomic and continuum scales
are tightly coupled \cite{Eldetal02,Eldetal04,Eldetal06,Steetal06,Beretal06}. 
This method is rooted in phenomenological continuum theories used to study 
equilibrium  and nonequilibrium patterns  
with ``crystal-like'' ordering in diverse contexts. The models most closely 
related to the PFC model in their mathematical formulation 
have appeared in studies of phase separation
in block copolymers \cite{FreHel87} and Rayleigh-B\'enard convection
\cite{SwiHoh77,CroHoh93}. 

Since the free-energy of the PFC model is a functional
of the density of the material, the model can also be cast \cite{Eldetal06} 
in the framework of classical density functional theory of freezing 
\cite{RamYus79,HayOxt81,Laietal87,HarOxt84,Sin91,SheOxt96a,SheOxt96b}. PFC 
simulations have the main advantage  of resolving  the atomic-scale density wave 
structure of a polycrystalline material and of describing the defect-mediated 
evolution of
this structure on time scales orders of magnitude 
longer than molecular dynamics (MD) simulations 
\cite{Eldetal02,Eldetal04,Eldetal06,Steetal06,Beretal06}.

While the PFC method has been shown to describe  
qualitatively a wide range of phenomena  
\cite{Eldetal02,Eldetal04,Eldetal06,Steetal06,Beretal06}, its
predictive capability in a crystal growth context 
remains largely unexplored. We investigate in this paper to what degree the
PFC model can reproduce quantitatively some key
equilibrium properties of the crystal-melt interface,
in particular the magnitude and anisotropy of the interfacial
free-energy $\gamma$. 
Well-developed atomistic methods to calculate these properties
\cite{broughton86,davidchack00,davidchack03,davidchack05,hoyt01,morris02}
have been applied to
both face-centered-cubic (fcc) \cite{hoyt01,hoyt02,morris02,sun04,sun041}
and body-centered-cubic (bcc) systems \cite{sun04,sun041,hoyt06,Wuetal06}.

Our study is based on 
the PFC model that is a reformulation of the
Swift-Hohenberg equation \cite{SwiHoh77} with
conserved dynamics introduced by Elder \emph{et al.} 
\cite{Eldetal02,Eldetal04}. This model favors 
bcc crystal ordering in three dimensions. Our analysis of 
this model is closely related 
to previous studies of melting carried
out in the framework of Ginzburg-Landau (GL) theory. The GL 
theory originally developed for bcc-liquid interfaces
by Shih {\it et al.} \cite{Shietal87}  
predates the PFC model and was recently revisited \cite{Wuetal06} in the
light of recent results from MD simulations. This re-examination 
showed that GL theory yields predictions of $\gamma$ and its
anisotropy in reasonably good agreement with MD simulations
for Fe, and the same MD results are used here to benchmark PFC model
predictions. 

GL theory is derived from classical density functional theory
that expresses the free-energy of the system as a functional
of its density distribution $n(\vec r)$, as in the PFC model.
Furthermore, it makes the strong assumption that $n(\vec r)$ can
be expanded as a sum 
\begin{equation}
n(\vec{\bf r}) = n_0 \left(1+\sum_{i} u_{i}(\vec r) \,\, e^{i\vec{K}_{i} \cdot 
\vec r}+\dots \right)\label{nrlv},
\end{equation}
of density waves corresponding to the
principal reciprocal lattice vectors (where 
the index $i$ spans the set of 12 \{$\vec K_{110}$\} 
vectors of the reciprocal fcc lattice for bcc ordering).
The amplitudes $u_i(\vec r)$ of these density waves
are the order parameters used to construct the GL free-energy. 
These amplitudes decay in the liquid at a rate that depends generally
on the angle between $\vec K_i$ and the
directional normal to the solid-liquid interface, which makes
$\gamma$ anisotropic. 
The fact that the anisotropy predicted by GL theory is in reasonably good
agreement with MD simulations suggests that this directional
dependence is a main determinant of 
anisotropy \cite{Wuetal06} .

Since the crystal density field of the PFC model is also dominated by the 
principal reciprocal lattice vectors, we expect this model to yield
similar predictions of bcc-liquid interfacial properties as GL theory. 
Of course, the two theories are not identical since the contribution of higher 
order reciprocal lattice vectors of magnitude larger than $|\vec K_{110}|$, 
corresponding to $``\dots"$ in Eq. (\ref{nrlv}) is small but non-vanishing in the 
PFC model. Furthermore, the strength of the nonlinear coupling 
between different density waves is determined by the form
of the free-energy functional in the PFC model, while it is determined in GL 
theory by using the simplifying assumption that all closed polygons composed
of principal reciprocal lattice vectors with the same 
number of sides have equal weight \cite{Shietal87,Wuetal06}.  
Despite these differences, we find here that the PFC model and GL theory yield 
very similar predictions of bcc-liquid interfacial properties
that are in reasonably good 
quantitative agreement with MD simulations. 

To relate formally the PFC model and GL theory, we carry out
a weakly-nonlinear multiscale analysis of the  PFC model.
This type of analysis, pioneered in the context of Rayleigh-B\'enard convection 
\cite{NewWhi69}, has provided a fundamental understanding of the universal 
behavior of nonequilibrium patterns close to the onset of instability 
\cite{CroHoh93}. It has also been revived recently in the framework of the 
renormalization group to derive computationally efficient implementations of the 
PFC model \cite{Goletal05,Athetal06}. In the pattern formation context where this 
analysis was first developed, the distance from the onset of instability can be 
characterized generally by a small parameter $\epsilon$, e.g. in 
Rayleigh-B\'enard convection $\epsilon \sim (R-R_c)/R_c$ where $R$ is the 
Rayleigh number and $R_c$ is its critical value corresponding to the onset of 
instability. Furthermore, close to onset ($\epsilon \ll 1$), spatially periodic 
patterns are generally slowly modulated in space. Considering the simplest case 
of a one-dimensional pattern for illustrative purposes, it is natural to write 
the field variables characterizing such a pattern in a form $\sim 
A(Z)e^{iq_0z}+c.c.$, where $Z\sim \epsilon^{1/2}z$ is a slow space variable, 
$q_0$ is the wavenumber of the perfectly ordered pattern, and $c.c.$ denotes the 
complex conjugate. The standard amplitude-equation approach  consists of using a 
multiscale expansion to obtain an equation for the complex amplitude $A(Z)$ 
starting from the underlying equations governing the evolution of the pattern. 
The complex amplitude $A(Z)\equiv u(Z)e^{i\Phi(Z)}$ carries information about 
both the local real
amplitude $u(Z)$ of the pattern and its local spatial periodicity, or 
wavenumber $q(Z)\approx q_0+\epsilon^{1/2}\partial_Z \Phi$.
Similarly, a dependence of the amplitude on a slow time variable (omitted from 
the present discussion) can also be introduced to describe the slow temporal 
evolution of the pattern.

For solid-liquid equilibrium, the pattern of interest is the three-dimensional 
crystal density field that is spatially modulated along the coordinate $z$ normal 
to the solid-liquid interface. However, there is no direct analog of a small 
parameter $\epsilon$ that can be made arbitrarily small by tuning some control 
parameter, such as the externally imposed temperature gradient in the example of 
Rayleigh-B\'enard convection. In contrast, $\epsilon$ is uniquely determined by 
liquid structure factor properties when relating the PFC model to classical DFT.
Thus $\epsilon$ has a fixed value for a given material. For systems with low 
entropy of melting and atomically rough interfaces, however, 
$\epsilon$ turns out to be small enough ($\sim 0.1$ for Fe) 
for a multiscale analysis to be just about justified quantitatively. This 
smallness originates physically from the fact that density waves decay slowly in 
the liquid over several atomic layer spacings. This makes $\epsilon$, which is 
proportional to the square of the ratio of the layer spacing and the interface 
width, much smaller than unity. For faceted interfaces, however, density waves 
decay abruptly in the liquid and this expansion would break down.

This paper is organized as follows. In section II, we briefly summarize the
equations of the PFC model and construct the phase-diagram corresponding to 
bcc-liquid coexistence. In section III, we derive the amplitude equations that 
describe the equilibrium profiles of density waves in the interface region
from the aforementioned multiscale expansion. The phases $\Phi$ of the complex 
amplitudes turn out to be constant in the interface region at dominant order in 
this expansion, such that the density field can be described by Eq. (\ref{nrlv}) 
with real order parameters that are the $u_i(\vec r)$'s. This allows us to define 
the free-energy as a functional of these order parameters and to compare in 
section IV the PFC amplitude equations to GL theory \cite{Wuetal06}. This 
comparison is used to fix uniquely the parameters of the bare PFC model in terms 
of liquid structure factor properties and the solid density wave amplitude 
derived from MD simulations. Differences in the nonlinear coupling between 
density waves in the PFC amplitude equations and GL theory are also highlighted 
in this section. In section V, we compare quantitatively the predictions of 
$\gamma$ for different crystal faces obtained using (i) the direct numerical 
solution of the PFC model, (ii) the amplitude equations derived from the PFC 
model, (iii) GL theory \cite{Wuetal06}, and (iv) MD simulations. Finally, 
concluding remarks are given in section V.   

\begin{figure}[t]
\includegraphics[width=0.5\textwidth]{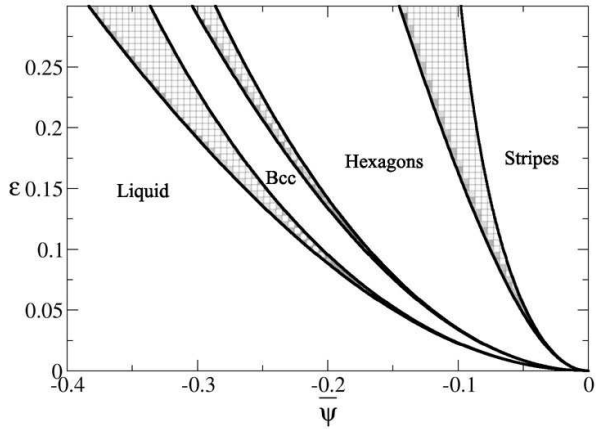}
\caption{Phase diagram of the PFC model obtained under the approximation that
the crystal density field is as a sum of density waves corresponding to the set 
of principal reciprocal lattice vectors for a given crystal structure.
}
\label{phase_diagram}
\end{figure}

\section{Phase-field crystal model}
\label{sec:pfc}

\subsection{Basic equations and scaling}
\label{sec:basic equations}

We consider the simplest PFC model defined by the free-energy
functional \cite{Eldetal02,Eldetal04}
\begin{equation}
\label{eq:pfc}
F = \int d\vec{\bf r} \left\{
{\phi \over 2}[
a + \lambda({q_0}^2+\nabla^2)^2
] \phi
+g{\phi^4 \over 4}  \right\},
\end{equation}
which is a transposition to crystalline solids of the
Swift-Hohenberg model of pattern formation \cite{SwiHoh77}. 
The conserved order parameter $\phi$ is
a dimensionless measure of the crystal 
density field measured from some constant reference value. 
The relationship of $\phi$  
to the physical density will
be specified in the next section.
The wavenumber $q_0$ sets the magnitude
$|\vec K_i|$ of the principal reciprocal lattice vectors 
that correspond to the first peak of the liquid structure factor
$S(K)$ at melting, and hence sets the scale 
of the ordered crystalline pattern $\sim q_0^{-1}$. 
As shown in the next section, the parameters $a$ and 
$\lambda$ can be related to properties of this peak and 
$g$, in turn, is uniquely fixed by the amplitude 
of density waves in the solid.

To render the calculations less cumbersome, it is useful
to rewrite the free-energy functional in dimensionless form
by defining the parameter
\begin{equation}
\label{epsdef}
\epsilon=-{a\over\lambda q_0^4},
\end{equation}
and making the substitutions,
\begin{equation}
\label{dimlessr}
q_0 \vec r \rightarrow \vec{r}, 
\end{equation}
\begin{equation}
\label{dimless}
\sqrt{g\over\lambda q_0^4} \phi \rightarrow \psi,
\end{equation}
\begin{equation}
{g\over \lambda^2 q_0^5} F \rightarrow {\cal F},\label{dimlessF}
\end{equation}
where all the transformed quantities to the right of the arrows 
are dimensionless and  
\begin{equation}
\label{eq:energy}
{\cal F} = \int d\vec r \left\{
{\psi \over 2}[
-\epsilon + (\nabla^2 + 1)^2
] \psi
+{1 \over 4} \psi^4  \right\}
\end{equation}
In this study, we restrict our attention to equilibrium properties of
the crystal melt interface. The condition that the 
chemical potential must be spatially uniform in equilibrium yields
the equation 
\begin{equation}
\mu_E={\delta {\cal F}\over 
\delta \psi} = 
  -\epsilon \psi +
(\nabla^2 +1)^2\psi +\psi^3, \label{eq:muE}
\end{equation} 
which is the starting point of the present study. Although the dimensionless 
formulation of the PFC model is more convenient to carry out calculations, we 
shall later transform the results back into dimensional form in order to make 
contact with GL theory and determine the phase-field parameters that appear in 
Eq. (\ref{eq:pfc}).
 
\subsection{Phase diagram}
\label{sec:phase}

To construct the phase diagram,   
we calculate separately the free-energy density 
(free-energy per unit volume) as a function of the mean density
$\bar \psi$ in solid, denoted by $f_s(\bar \psi)$, 
and liquid, $f_l(\bar \psi)$, using Eq. (\ref{eq:energy}). 
We then use the standard common 
tangent construction, which is equivalent to equating the chemical
potentials and grand potentials of the two phases, to obtain
the equilibrium values of $\bar \psi$ in the
solid ($\bar \psi_s$) and liquid ($\bar \psi_l$).

Since the density is constant in the liquid, $f_l$ 
is obtained directly from
Eq. (\ref{eq:energy})    
\begin{equation}
f_l=-(\epsilon-1)\frac{\bar\psi^2}{2}+\frac{\bar\psi^4}{4}
\label{fldef}
\end{equation}
Furthermore, since $\epsilon$ turns 
out to be a small parameter for spatially diffuse atomically
rough interfaces, the solid free-energy density 
can be well approximated by only considering the contribution
of the principal reciprocal
lattice vectors. Accordingly, 
the crystal density field can be
written in the form analogous to Eq. (\ref{nrlv}) 
\begin{eqnarray}
\label{eq:bccs}
\psi(\vec{r})&\approx& \bar\psi  + \sum_i A_i\, e^{i\vec{K_i}\cdot \vec{r}} \\ 
\nonumber
&\approx& \bar\psi +  \\ \nonumber 
&&4A_s\,(\cos{q x} \cos{q y}  
+\cos{q x} \cos{q z} +  
\cos{q y} \cos{q z}),
\end{eqnarray}
where we have used the fact that 
all density waves have the same amplitude ($|A_i|=A_s$)
and all principal reciprocal lattice
vectors of the bcc structure have
the same magnitude  
($|\vec K_{110}|=|\vec K_{1-10}|=...=\sqrt{2}q$). 
The parameters $A_s$ and $q$ are solved by 
substituting Eq.~(\ref{eq:bccs}) into Eq.~(\ref{eq:energy}) and 
minimizing the resulting free-energy with
respect to $A_s$ and $q$, which yields
\begin{equation}
A_s=-{2 \over 15}\bar\psi+{1 \over 15}
\sqrt{5\epsilon-11\bar\psi^2}\label{Asdef}
\end{equation} 
and $q=1/\sqrt{2}$, together with the expression
for the solid free-energy density (for $q=1/\sqrt{2}$) 
\begin{eqnarray}
f_s&=&-(\epsilon-1)\frac{\bar\psi^2}{2}+\frac{\bar\psi^4}{4}
\nonumber\\
& &-6\epsilon A_s^2+18\bar\psi^2A_s^2+48\bar\psi A_s^3+135 A_s^4.\label{fsdef}
\end{eqnarray}
Applying the common tangent construction, which is detailed below in
the small $\epsilon$ limit, yields the bcc-liquid coexistence region in the phase 
diagram of Fig. \ref{phase_diagram}. 
Also shown are the other two-dimensional 
crystal structures (hexagonal and stripe phases) determined in previous studies 
using the same approximation where the crystal density field is a sum of
density waves corresponding to the set of principal reciprocal lattice vectors 
\cite{Eldetal02,Eldetal04}.

\section{Derivation of the amplitude equations}
\label{sec:III}

\subsection{Small $\epsilon$ analysis of the phase diagram}

For small $\epsilon$, we can seek  
a perturbative solution of the crystal density 
field $\psi$ of the form
\begin{equation}
\label{eq:expansion}
\psi(\vec{r})=\psi_0(\vec{r})\,\epsilon^{1/2} + \psi_1(\vec{r})\,\epsilon + 
\psi_2(\vec{r})\,\epsilon^{3/2}+\dots,
\end{equation}
and expand accordingly the average densities 
\begin{equation}
\label{psis}
\bar{\psi_s} = \psi^0_s\, \epsilon^{1/2}+\psi^1_s\, \epsilon+\psi^2_s\, 
\epsilon^{3/2}+ \dots,
\end{equation}
and
\begin{equation}
\label{psil}
\bar{\psi_l} = \psi^0_l\, \epsilon^{1/2}+\psi^1_l\, \epsilon+\psi^2_l\, 
\epsilon^{3/2}+ \dots,
\end{equation}
in the solid and
liquid, respectively.
Substituting these relations into the expressions
for $f_s$ and $f_l$, using the conditions of
equality of the chemical potentials of the two phases,
$f_s'(\bar\psi_s)=f_l'(\bar\psi_l)=\mu_E$, 
and equality of the grand potentials
$f_s(\bar\psi_s)-\mu_E \bar\psi_s=f_l(\bar\psi_l)-\mu_E \bar\psi_l$,
and collecting powers of $\epsilon$, we obtain  
\begin{equation}
\psi^0_s=\psi^0_l\equiv \psi_c=-\sqrt{45\over 103},
\end{equation}
and 
\begin{equation}
\psi^1_s=\psi^1_l=0.
\end{equation}
This shows that, in the small $\epsilon$ limit, the PFC
model exhibits a weak first-order freezing transition where
the size of the solid-liquid coexistence region   
$\Delta \bar{\psi}=\bar{\psi_s}-\bar{\psi_l} \approx (\psi^2_s-\psi^2_l)
\,\epsilon^{3/2}$ is much smaller than the mean value of the density $\sim 
\epsilon^{1/2}$. These scalings imply that the mean density 
difference between the two phases only gives a small higher order
correction to the density wave profiles through the
interface and $\gamma$ in the small $\epsilon$ limit.

\subsection{Multiscale expansion}

Using Eqs. (\ref{psis}) and (\ref{psil}) to evaluate the small $\epsilon$
limit of the chemical potential $\mu_E=f_l'(\bar\psi_l)=f_s'(\bar\psi_s)$,
the equilibrium equation of the density field (\ref{eq:muE}) becomes 
\begin{eqnarray}
\label{main}
& & -\epsilon \psi + (\nabla^2+1)^2\psi +{\psi}^3 \nonumber \\
&=& \psi_c \epsilon^{1/2} + (\psi_l^2- \psi_c +{\psi_c}^3)\epsilon^{3/2}
+\dots
\end{eqnarray}
The derivation of the amplitude equation exploits the separation of
scale between the width of the spatially diffuse interface and the interatomic
layer spacing in the small $\epsilon$ limit. This separation
of scale allows us to assume that the envelope of density waves depends on a slow 
spatial variable $Z \equiv \epsilon^{1/2}\,z$ (i.e. $\psi_0(\vec{r})=\psi_c+\sum
A_i^0(Z) e^{i\vec{K_i}\cdot \vec{r}}$, and so on for higher order terms)
where $z$ denotes the coordinate along the direction normal to the solid-liquid 
interface. The multiscale expansion rests on treating the slow variable $Z$
and the fast variable $z$ as independent variables. Thus the spatial
derivative along $z$ transforms with the chain rule  
 $\partial_z \rightarrow \partial_z + \epsilon^{1/2} \partial_Z$,
and the differential operator $L^2 \equiv (\nabla^2+1)^2$
in Eq. (\ref{main}) becomes
\begin{equation}
L^2  \rightarrow L^2 + 
4\epsilon^{1/2} L \partial_z \,\partial_Z 
+2\epsilon (L+2\partial_z^2)\,\partial_Z^2,
\end{equation}
where the differential operator
$L$ on the right-hand-side   
only acts on the fast spatial variable $z$.

Next, we substitute the small $\epsilon$ expansion
of the density field (\ref{eq:expansion}) into 
the equilibrium equation (\ref{main}) with the above
transformation of the linear operator.  
Collecting terms with the same power $\epsilon$, we find
at the order $\epsilon^{1/2}$
\begin{equation}
{L}^2 \psi_0 = \psi_c,
\end{equation}
which has the solution 
\begin{eqnarray}
\psi_0 = \sum_i A_i^0(Z)  e^{i \vec{K_i} \cdot \vec{r}} +\psi_c, 
\end{eqnarray} 
where $|\vec K_i|=1$ in our scaled units.
At order $\epsilon$, we obtain 
\begin{equation}
{L}^2 \psi_1 =0,
\end{equation}
which has the solution 
\begin{eqnarray}
\psi_1 = \sum_i A_i^1(Z)  e^{i \vec{K_i} \cdot \vec{r}},  
\end{eqnarray} 
and collecting the terms at  
order $\epsilon^{3/2}$ yields 
\begin{equation}
\label{eq:third}
{L}^2 \psi_2 + (4\,\partial_z^2\,
\partial_Z^2-1)\psi_0 + {\psi_0}^3 = \psi_l^2 -\psi_c + {\psi_c}^3.
\end{equation} 
The amplitude equations are obtained from the condition
for the existence of a solution of the above equation without
needing to compute $\psi_2$ explicitly. 
Since ${L}^2 \psi_2$ gives a vanishing contribution
for all density waves associated with the
set $\{\vec K_i\}$ of twelve principal reciprocal lattice vectors
of magnitude unity
(i.e., ${L}^2 e^{i \vec{K} \cdot \vec{r}}
=(-|K|^2+1)^2e^{i \vec{K} \cdot \vec{r}}=0$ if $|\vec K|^2=1$), all
remaining terms $\sim e^{i\vec K_i\cdot \vec r}$ must
balance each other in order for a solution of 
Eq. (\ref{eq:third}) to exist.  
For example, the condition that the
coefficients of $e^{i \vec{K_{011}} \cdot \vec{r}}$
balance each other, yields
\begin{eqnarray}
\label{ampeqn011}
&&(4(\hat{K_{011}} \cdot \hat{n})^2\,\partial^2_Z+3{\psi_c}^2-1)A^0_{011} +
\left(  3|A^0_{011}|^2 + 6|A^0_{110}|^2  \right. \nonumber \\
&&+ 6|A^0_{1\bar{1}0}|^2
+ 6|A^0_{101}|^2
 + 6|A^0_{10\bar{1}}|^2 +
\left. 6|A^0_{01\bar{1}}|^2 \right)   A^0_{011} 
\nonumber \\
&&+6 A^0_{01\bar{1}}A^{0*}_{10\bar{1}}A^0_{101} + 6 
A^{0*}_{01\bar{1}}A^0_{110}A^{0*}_{1\bar{1}0}
\nonumber \\
&& +6A_{101}^0A^{0*}_{1\bar{1}0}{\psi}_c + 6A^0_{110}A^{0*}_{10\bar{1}} {\psi}_c
=0,
\end{eqnarray}
where everywhere in this paper $\hat{z}=\hat{n}$ corresponds to the direction
normal of the interface that generally
differs from the crystal axes except for $\{100\}$ crystal faces.  
This solvability condition
must be satisfied independently for each $\vec K_i$. This yields
a set of twelve coupled amplitude 
equations (i.e., eleven additional equations to the one
above) that are straightforward to obtain and we do not list them all
here for brevity of presentation. These equations can also be obtained
directly from the free-energy expressed as a functional of the  
amplitudes $A_i^0(Z)$ as described in the next subsection.

The amplitude profiles are governed 
by these twelve coupled nonlinear 
amplitude equations. 
These equations can be reduced to a simple set of equations by considering
the symmetry of reciprocal lattice vectors. 
For $\{100\}$ crystal faces, 
these twelve amplitudes can be separated into two subsets with the same
value of $(\vec{K_i} \cdot \hat{n})^2$ equals to $1/2$ and $0$
respectively.
Therefore, the amplitude equations are reduced to only two coupled
equations and can be solved numerically.
Similarly, we have two subsets of amplitudes for $\{111\}$ crystal
faces and three subsets of amplitudes for $\{110\}$ crystal faces, which results 
in two and three coupled amplitude equations, respectively.

As in GL theory \cite{Wuetal06},
the $\gamma$ anisotropy originates from the
fact that the coefficients of the second
derivative terms in the amplitude equations
depend on $(\hat{K_i} \cdot \hat{n})$ and hence on the orientation
of the crystal face with respect to a fixed set of crystal axes.
Furthermore, as mentioned in the introduction, since the amplitudes  
are complex, the spatial variation of the phase  
can cause the local wave vector to change through the
solid-liquid interface by an amount 
proportional to the gradient of this phase. To determine
this variation, we substitute 
\begin{equation}
A_{i}^0 (Z) = |A_i^0 (Z)|  e^{i\Phi_i(Z)}
\end{equation}
into the amplitude equations. We obtain that
$\Phi_i(Z)=0$ for the principal reciprocal lattice vectors 
that are orthogonal to the interface normal, and  
\begin{equation}
{1\over |A_i^0(Z)|} \partial_Z (|A_i^0(Z)|^2 \Phi_i(Z)) = 0
\end{equation}
for the other reciprocal lattice vectors.
The above equation implies that 
\begin{equation}
|A_i^0(Z)|^2 \partial_Z \Phi_i(Z)=C_0
\end{equation}
where $C_0$ is a constant.
Since the amplitudes must vanish in liquid, the divergence of $d \Phi_i/dZ$
can be avoided only if $C_0=0$. Therefore, the  
wave vectors ${\vec K_i}$'s are constant through the solid-liquid interface 
in the small $\epsilon$ limit.

\subsection{Free-energy functional}

It is useful to express the free-energy of the solid-liquid
system as a functional of the density wave amplitudes $A_i^0$. 
For this, we define $\Delta {\cal F}$ to be the free-energy measured from
its constant value in the liquid. Since the amplitudes
are non-conserved order parameters, the equilibrium state simply
corresponds to a minimum of this free-energy without extra
constraint. This implies that
$\Delta {\cal F}$ should be chosen such that the  
amplitude equations are recovered
variationally from this free-energy. Namely,  
the equation for a given
$A_i^0$ derived in the last subsection should be
equivalent to
\begin{equation}
\label{eq:potential}
\frac{\Delta {\cal F}}{\delta A_i^{0*}} =0,  
\end{equation}
up to a multiplicative constant.
This constant can be determined by matching the
limiting value of $\Delta {\cal F}/V$ on the solid-side, where all
the amplitudes are constant ($A_i^0 = \epsilon^{-1/2} A_s$ for all $i$) 
to the difference of free-energy densities
between the two phases, 
$f_s-f_l$, where $f_s$ and $f_l$ are given
by Eqs. (\ref{fldef}) and (\ref{fsdef}) and $V$ is the volume. 
This yields the free-energy functional,
\begin{equation}
\label{eq:amp_pot}
{\Delta {\cal F}} = \epsilon^{3/2}  \Omega \int dZ \,\left[ \sum_{i} \,2 
(\hat{K_i}\cdot \hat{n})^2 \,\left|{dA_i^0\over dZ}\right|^2 + f({A_i^0})\right],
\end{equation}
where $\Omega\equiv \int dx dy$ is the interface area and
\begin{eqnarray}
& &f({A_i^0}) =
{1\over 2}\sum_{i} (3{\psi_c}^2-1){|A_i^0|}^2 
+{3\over 4}\sum_{i}\sum_{j\neq i}\,{|A_i^0|}^2{|A_{j}^0|}^2
\nonumber \\
& &+6{A_{110}^0}^*{A_{1\bar{1}0}^0}^*{A_{101}^0}{A_{10\bar{1}}^0}
+6{A_{110}^0}{A_{1\bar{1}0}^0}{A_{101}^0}^*{A_{10\bar{1}}^0}^*
\nonumber \\
& &+6{A_{1\bar{1}0}^0}{A_{011}^0}{A_{01\bar{1}}^0}{A_{110}^0}^*
+6{A_{1\bar{1}0}^0}^*{A_{011}^0}^*{A_{01\bar{1}}^0}^*{A_{110}^0}
\nonumber \\
& &+6{A_{01\bar{1}}^0}{A_{10\bar{1}}^0}^*{A_{101}^0}{A_{011}^0}^*
+6{A_{01\bar{1}}^0}^*{A_{10\bar{1}}^0}{A_{101}^0}^*{A_{011}^0}
\nonumber \\
& &+6\psi_c {A_{011}^0}^* {A_{101}^0} {A_{1\bar{1}0}^0}^*
+6\psi_c {A_{011}^0} {A_{101}^0}^* {A_{1\bar{1}0}^0}
\nonumber \\
& &+6\psi_c {A_{011}^0}^* {A_{110}^0} {A_{10\bar{1}}^0}^*
+6\psi_c {A_{011}^0} {A_{110}^0}^* {A_{10\bar{1}}^0}
\nonumber \\
& &+6\psi_c {A_{01\bar{1}}^0}^* {A_{110}^0} {A_{101}^0}^*
+6\psi_c {A_{01\bar{1}}^0} {A_{110}^0}^* {A_{101}^0}
\nonumber \\
& &+6\psi_c {A_{01\bar{1}}^0}^* {A_{10\bar{1}}^0} {A_{1\bar{1}0}^0}^*
+6\psi_c {A_{01\bar{1}}^0} {A_{10\bar{1}}^0}^* {A_{1\bar{1}0}^0}\nonumber\\
\end{eqnarray}
It is simple to check that by applying Eq. (\ref{eq:potential}) 
to the above functional for $A_i^0=A_{011}$ we obtain 
the same amplitude equation as Eq. (\ref{ampeqn011}) and similarly
for the other principal reciprocal lattice vectors. Finally, Eq.
(\ref{dimlessF}) implies that the dimensional free-energy functional
derived from the amplitude equations (AE) is given by
\begin{equation}
\Delta F_{AE}={\lambda^2 q_0^5 \over g}\Delta  {\cal F}. \label{scaleback}
\end{equation}

\section{Comparison of Amplitude Equations and 
Ginzburg-Landau theory}

In this section, we compare the free-energy functional derived
from the PFC amplitude equations to GL theory \cite{Wuetal06}. 
This comparison sheds light on the relation between this theory and the
PFC model and uniquely fixes the parameters of the latter in terms
of physical quantities that can be extracted from MD simulations. 

\subsection{Ginzburg-Landau theory}

The free-energy functional of GL theory is expressed in terms
of the amplitude $u_i$ of density waves defined by Eq. (\ref{nrlv}).
We write this functional here for convenience 
using the same notation as
in Ref. \cite{Wuetal06} 
\begin{eqnarray}
\label{GLF}
\Delta F_{GL}&=& {n_0 k_B T\Omega\over2} \left(\int dz \,
a_2 \sum_{i, j} c_{ij}\,u_{i}\, u_{j}
\,\,\delta_{0,\vec{K}_i+\vec{K}_j}  \right.  \nonumber \\
&-a_3& \sum_{i,j,k} c_{ijk}\, u_{i} \, u_{j} \, u_{k}\,\, 
\delta_{0,\vec{K}_i+\vec{K}_j+\vec{K}_k} \\ \nonumber
&+a_4& \sum_{i,j,k,l} c_{ijkl} \,u_{i} \, u_{j}\, u_{k}\, u_{l}
\,\,\delta_{0,\vec{K}_i+\vec{K}_j+\vec{K}_k+\vec{K}_l} \\ \nonumber
&+ b& \left. \sum_{i} c_{i}\,\left|{du_{i}\over dz}\right|^2 \right),
\end{eqnarray}
where $\delta_{m,n}$ is the Kronecker delta
that equals $0$ or $1$ for $m\ne n$ or $m= n$, respectively. 
The latter enforces that only combinations of principal reciprocal lattice
vectors that form closed polygons $\vec{K}_i+\vec{K}_j+\dots=0$ contribute to
the free-energy functional. The multiplicative factors
$a_i$ and $b$ are introduced
since it is convenient to normalize
the sums of the $c$'s to unity (i.e.
$\sum_{i} c_{i} = 1$, $\sum_{i,j}c_{ij}\delta_{0,\vec{K}_i+\vec{K}_j}=1$, etc).

The coefficients of quadratic nonlinearities 
of the GL free-energy were determined in
Ref. \cite{Wuetal06} by relating $\Delta F_{GL}$ to 
the free-energy functional  
that describes small density fluctuations of an inhomogeneous liquid 
in the simplest formulation of DFT
(Eq. (3) in Ref. \cite{Wuetal06}). In particular,
the latter can be reduced to the form \cite{HayOxt81}
\begin{eqnarray}
\label{DFT2}
\Delta F_{DFT} &\approx&  {n_0 k_B T \Omega\over 2} \int dz
 \left[ \sum_{i,j}{1\over  S(|\vec K_i|)} u_iu_j \delta_{0,\vec{K}_i+\vec{K}_j}
\right. \nonumber \\
& &\left.
-\sum_i {1\over 2} C''(|\vec K_i|)
(\hat{K}_{i} \cdot \hat{n})^2 \left|{du_{i}\over dz}\right|^2 \right]
\end{eqnarray}
by assuming that the density wave amplitudes vary slowly through
the interface region and are essentially
constant on the scale of the interatomic layer spacing. 
The small $\epsilon$ multiscale expansion of the last
section is an alternative procedure to derive the form 
of Eq. (\ref{DFT2}) that formalizes this assumption.
Here $C(K)$ is the 
fourier transform of the direct correlation function $C(|\vec{\bf r}|)$
\begin{equation}
C(K)=n_0 \int d\vec {\bf r}\, C(|\vec{\bf r}|) e^{-i \vec{K} \cdot \vec{\bf r}},
\end{equation}
and $S(K)=\left[1-C(K)\right]^{-1}$ is the liquid structure factor.    

Equating $\Delta F_{GL}$ and $\Delta F_{DFT}$ at quadratic
order in the nonlinearities and using the normalization
that the sums of $c_i$'s and $c_{ij}$'s equal unity, we obtain
\begin{equation}
c_{ij}=1/12,
\end{equation}
\begin{equation}
c_i={1\over4}(\hat{K}\cdot\hat{n})^2,
\end{equation}
\begin{equation}
a_2 = {12 \over S(K_{max})},\label{a2GL}  
\end{equation}
\begin{equation}
b = -2 C''(K_{max}), \label{bGL} 
\end{equation}
where the magnitude $|\vec K_i|=q_0$ of the 
principal reciprocal lattice vectors can be set
equal to the $K$ value corresponding 
to the first peak of the structure factor, $K_{max}$, under the
assumption that the wave vectors are constant in the
interface region. This assumption was formally justified in the
derivation of the amplitude equations in Sec. III B by
showing that the phase $\Phi$ of the complex amplitudes is 
constant in the interface region
at leading order in the small $\epsilon$ expansion. 

The reader is referred to Ref. \cite{Wuetal06} for the
determination of the cubic and quartic
nonlinearities in the GL theory, which shall be briefly
reviewed below. 

\subsection{Determination of phase-field crystal
model parameters}

We are now in a position to compare 
the free-energy functionals derived from the amplitude equations
and GL theory and to relate the parameters 
of the PFC model to physical quantities. 
For this, we note that $\Delta F_{AE}$
has the same form as $\Delta F_{GL} $
because the density wave amplitudes 
$A_i^0$'s and $u_i$'s
proportionally related.
The proportionality constant
is readily obtained by combining Eq. (\ref{dimless}) and Eq. 
(\ref{eq:expansion}),
which yields
\begin{equation}
\label{uiAi}
n_0 u_i = \sqrt{\lambda q_0^4\over g} \epsilon^{1/2}A_i^0.
\end{equation}
Furthermore, Eq. (\ref{eq:potential}) used to
construct $\Delta F_{AE}$ is equivalent to the constraint 
that only combinations of principal reciprocal lattice
vectors that form closed polygons contribute to the
free-energy functional.   

Next, using Eq. (\ref{uiAi})  
and equating $\Delta F_{AE}$, defined by Eqs. 
(\ref{eq:amp_pot}) and (\ref{scaleback}), and 
$\Delta F_{GL}$ defined by Eq. (\ref{GLF}), we
obtain the relations  
 \begin{equation}
a_2 = {{12\,n_0\,a(1-3{\psi_c}^2)}\over k_B T}={12 \over S(K_{max})},\label{afix}
\end{equation}
\begin{equation}
b = {{16\, n_0 \,\lambda q_0^2}\over {k_B T}}=-2 C''(K_{max}),\label{lambdafix}
\end{equation}
where we made use of Eqs. (\ref{a2GL}) and (\ref{bGL}) to
write the second equalities and
$\psi_c=-\sqrt{45/103}$ as shown earlier. 
Eqs. (\ref{afix}) and(\ref{lambdafix})  
uniquely relate the parameters $a$ and $\lambda$ of the PFC model to peak 
properties of the liquid structure factor that can be computed from MD 
simulations
or measured experimentally. They also fix the value of $\epsilon$
related to $a$ and $\lambda$ by Eq. (\ref{epsdef})
\begin{equation}
\epsilon = {8 \over (1-3{\psi_c}^2)\, q_0^2 \, S(K_{max}) \, C''(K_{max})}.
\end{equation}
The only left unknown parameter $g$ of the PFC model can be obtained 
by applying Eq. (\ref{uiAi}) in the solid 
where all the density wave amplitudes have equal magnitude.
Substituting into Eq. (\ref{uiAi}) $u_i=u_s$ and the solid
value of the amplitudes 
\begin{equation}
\label{As}
A_i^0 =\epsilon^{-1/2}A_s= -{2 \over 15} \psi_c + {1 \over 15} \sqrt{5-11 
{\psi_c}^2},
\end{equation}
which follows from Eq. (\ref{Asdef}) 
or Eq. (\ref{ampeqn011}), we obtain the relation
\begin{equation}
g=\lambda q_0^4\left(-{2 \over 15} \psi_c + {1 \over 15} \sqrt{5-11 
{\psi_c}^2}\right)^2/(n_0^2u_s^2)
\end{equation}
This relation fixes $g$ in terms 
of the other parameters and $u_s$, which 
can be extracted directly from MD simulations \cite{Wuetal06} or related
to the latent heat of melting and the 
temperature dependence of $S(K_{max})$ 
\cite{Shietal87,Wuetal06}.
 
\subsection{Coefficients of quartic nonlinearities}
\label{subsec:quartic}
 
The free-energy functionals derived from
the PFC amplitude equations and GL theory only differ in
the values of the coefficients of higher order
nonlinearities. As we shall see in the next section, these
differences turn out to be unimportant because the amplitude
equations and GL theory yield essentially identical predictions
of $\gamma$ and its anisotropy. However, they deserve
brief mention.
In GL theory, the coefficients $a_3$ and $a_4$  
are determined from the two equilibrium conditions that 
(i) the solid and liquid
phases must have equal free-energies at melting, $\Delta F_{GL}|_{u_i=u_s}=0$, 
and (ii) the equilibrium state 
of the solid is a minimum of free-energy, $\partial
\Delta F_{GL} / \partial u_i |_{u_i=u_s}$=0. These two conditions
 yields the relations \cite{Shietal87,Wuetal06} 
\begin{equation}
 a_3=2a_2/u_s,
\end{equation}
 and
\begin{equation}
 a_4=a_2/u_s^2.
\end{equation}
The amplitude-equation free-energy functional   
$\Delta F_{AE}$ satisfies automatically 
the above two equilibrium conditions by construction. 
Thus, it only differs from $\Delta F_{GL}$ in the
calculation of the other 
coefficients of the cubic and quartic terms, $c_{ijk}$ and
$c_{ijkl}$. In GL theory, these coefficients   
by the ansatz that all closed polygons  of
$\vec K_i$'s with the same number of sides have 
the same weight, which yields
$c_{ijk}=1/8$ and $c_{ijkl}=1/27$ \cite{Wuetal06}.
 
In contrast, in the PFC amplitude equations, these coefficients
are uniquely determined by the choice of the nonlinear
terms in the original PFC free-energy functional. 
For the simplest choice of nonlinearity $\sim \phi^4$ considered
here, the amplitude equation derivation yielded  
the same coefficients of cubic terms as GL theory but different
coefficients of quartic terms. 
Comparing Eq. (\ref{GLF}) with Eqs. (\ref{eq:amp_pot}) 
and (\ref{scaleback}), we obtain that, in the expression
for $\Delta F_{AE}$, 
$c_{ijkl}=1/90$ for two-sided polygons that contain only two wave vectors 
$\vec K_i$ and $-\vec K_i$, and $c_{ijkl}=4/90$ for the rest 
of the quartic terms.

To make these differences explicit, we consider the  
$\{110\}$ crystal faces. The set of 12 principal reciprocal lattice vectors
corresponding to $\langle 110 \rangle$ 
direction can be separated into three subsets
with the same value of $(\hat{K_i} \cdot \hat{n})^2$:
subset I with 8 vectors
($[011],[0\bar 11],[01\bar 1],[101],
[\bar 1 01],[10\bar 1],[0\bar 1\bar 1],[\bar 10\bar 1]$)
and $(\hat K_i\cdot \hat n)^2=1/4$, subset
II with 2 vectors ($[110],[\bar 1\bar 10]$) and
$(\hat K_i\cdot \hat n)^2=1$, and subset III with 2 vectors
($[\bar 110]$, $[1\bar 10]$) and $(\hat K_i\cdot \hat n)^2=0$.
Density waves
in a given subset have the same amplitude denoted
here by  $u$, $v$, and $w$ for subsets
I, II and III, respectively. 
Then for $\{110\}$ crystal faces, Eqs. (\ref{eq:amp_pot}) 
and (\ref{scaleback}) reduce to
\begin{eqnarray}
\label{eq:PFCenergy}
{\Delta F}_{AE} &=& {n_0 k_B T\Omega \over 2} \int dz \left[
a_2\left({2\over 3} u^2 + {1\over 6} v^2 + {1\over 6} w^2
\right)\right. \nonumber \\
&-a_3&\left({1\over 2} u^2 v + {1\over 2} u^2 w\right)
+a_4\left({36\over 90} u^4 + {1\over 90} v^4 +{1\over 90} w^4\right. 
\nonumber \\
& &\left. +{16\over 90} u^2 v^2
+{16\over 90} u^2 w^2 +{4\over 90} w^2 v^2 +{16\over 90} u^2 v w\right) \nonumber
\\
& &\left. -C''(|\vec K_{110}|) \left|{du\over dz}\right|^2
-C''(|\vec K_{110}|)\left|{dv\over dz}\right|^2\right],
\end{eqnarray}
and differ from the corresponding expression obtained
from GL theory  
\begin{eqnarray}
\label{eq:GLenergy}
{\Delta F}_{GL} &=& {n_0 k_B T \Omega\over 2} \int dz \left[
a_2\left({2\over 3} u^2 + {1\over 6} v^2 + {1\over 6} w^2
\right)\right. \nonumber \\
&-a_3&\left({1\over 2} u^2 v + {1\over 2} u^2 w\right)
+a_4\left({12\over 27} u^4 + {1\over 27} v^4 +{1\over 27} w^4\right. 
\nonumber \\
& &\left. +{4\over 27} u^2 v^2
+{4\over 27} u^2 w^2 +{1\over 27} w^2 v^2 +{4\over 27} u^2 v w\right) \nonumber
\\
& &\left. -C''(|\vec K_{110}|) \left|{du\over dz}\right|^2
-C''(|\vec K_{110}|)\left|{dv\over dz}\right|^2\right].
\end{eqnarray}

\section{Comparison of continuum theories and molecular dynamics simulations}
\label{sec:IV}

In Ref. \cite{Wuetal06},
the predictions of GL theory were compared to
MD simulations of Fe with interatomic potentials
developed by Mendelev,
Han, Srolovitz, Ackland, Sun, and Asta (MH(SA)$^2$) based
on the embedded atom method \cite{MHSA2}.
In this section, we extend this comparison to include the
predictions of both the PFC model, with the free-energy functional
defined by Eq. (\ref{eq:pfc}) and the amplitude equations
derived from this model with the free-energy functional
defined by Eqs. (\ref{eq:amp_pot}) 
and (\ref{scaleback}). We use the same MD simulation
results for the present comparison.  
Details of the MD simulations  
and of the method to extract the density wave profiles from
these simulations are given in Ref. \cite{Wuetal06} and need not be 
repeated here.

The input parameters for the different
continuum theories are computed from
the MD simulations in order to make
the comparison with these simulations
as quantitative and precise as possible.
These include the parameters related to peak properties of the liquid
structure factor $K_{max}=q_0=2.985\,\,{\AA}^{-1}$, $1/S(K_{max}) = 0.332$ ,
$C''(K_{max})=-10.40 \,\,{\AA}^2$, and the amplitude of density waves 
corresponding to the principal reciprocal
lattice vectors in the solid $u_s=0.72$. These input parameters
fix the various coefficients of the continuum theories derived
in the last section, which 
are listed in Table \ref{tabparam}.
 
\begin{table*}[h]
\caption{Values of input parameters  
from MD simulations with interatomic EAM potential for Fe from MH(SA)$^2$ 
\cite{MHSA2} and resulting coefficients used in GL theory, the PFC model, and the
amplitude equations derived from this model.} 
\centering
\begin{tabular*}{0.9\textwidth}%
     {@{\extracolsep{\fill}}cccccccccc} \hline
 &$n_0$ ($\AA^{-3}$)&  $a_2$ &$b$ ($\AA^2$) & $u_s$ & $q_0$ ($\AA^{-1}$) & $a$ (eV $\AA^3$)& 
$\lambda$ (eV $\AA^7$) & $g$ (eV $\AA^9$)  &$\epsilon$ \\ \hline
MD (MH(SA)$^2$) (Ref. \cite{sun04})& 0.0765&3.99 & 20.81 & 0.72 & 2.985 & $-$2.136 & 
0.291 & 9.705 &0.0923 \\ \hline
\end{tabular*}
\label{tabparam}
\end{table*}

The calculation of density wave profiles and
$\gamma$ values for the PFC amplitude equations,
Eqs. (\ref{eq:amp_pot}) and (\ref{scaleback}),
proceeds in the same way as for GL theory \cite{Wuetal06}.
For example, for the case of the $\{110\}$ crystal faces
elaborated in section \ref{subsec:quartic},
the density wave profiles were
calculated by minimizing  $\Delta F_{AE}$  
given by Eq. (\ref{eq:PFCenergy}) with respect
to the order parameters $u$, $v$ and $w$,
and by solving numerically the resulting set
of coupled ordinary differential equations with
the boundary condition $u=v=w=u_s$ in solid
and $u=v=w=0$ in liquid. The value of 
$\gamma_{110}=\Delta F_{AE}/\Omega$ was then computed by integration
of Eq. (\ref{eq:PFCenergy}) with these profiles.
The same procedure was repeated for the $\{100\}$
and $\{111\}$ crystal faces, with different set of order
parameters for each crystal face.

To compute $\gamma$ in PFC simulations, we first
relax the density field $\psi$ to a minimum of the
free-energy functional ${\cal F}\equiv \int d\vec r f$,
where the free-energy density
$f$ is the integrand of Eq. (\ref{eq:energy}), using
a simple diffusive dynamics. We then compute $\gamma$ 
in dimensional units using the relation
\begin{equation}
\gamma=\Omega^{-1}\frac{\lambda^2 q_0^5}{g}\int d\vec r
\left[f-\left(f_s\frac{\psi-\bar \psi_l}{\bar \psi_s-\bar 
\psi_l}-f_l\frac{\psi-\bar \psi_s}{\bar \psi_s-\bar \psi_l}\right)\right]
\end{equation}
where  $\bar \psi_s$ ($\bar \psi_l$) and $f_s$ ($f_l$) are the mean values
of $\psi$ and the free-energy density in solid (liquid), respectively,
and $\Omega=\int dxdy$ is the interface area.
Although $\epsilon$ is small, these values 
need to be computed numerically (i.e. by calculating the solid
free-energy density from the numerical solution of the PFC model
rather than using the weakly nonlinear approximations derived
in section II) in order to obtain an accurate computation of $\gamma$.

\begin{figure}[t]
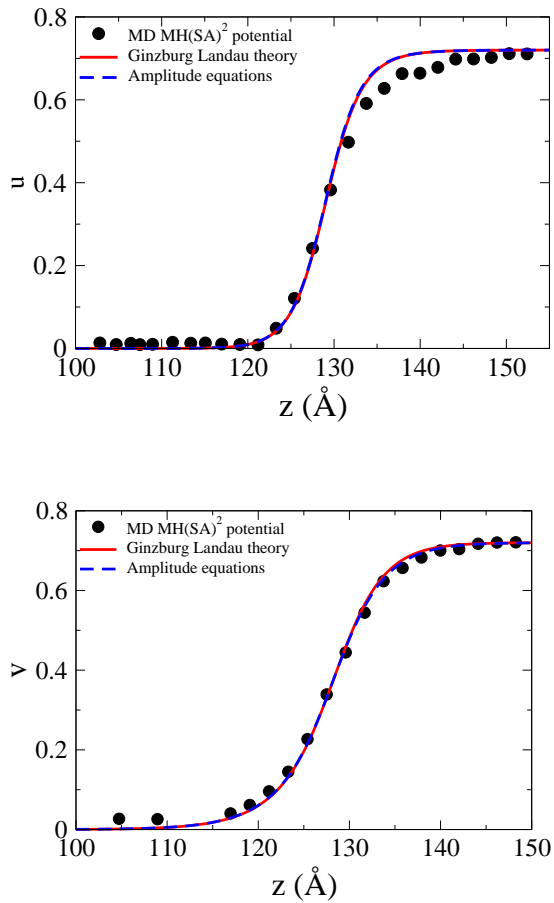

\begin{minipage}[t]{8cm}
\includegraphics[width=0.9\textwidth]{w_101_pfc_paper.eps}
\end{minipage}
\hfill
\vskip 1cm
\begin{minipage}[t]{8cm}
\includegraphics[width=0.9\textwidth]{w_110_pfc_paper.eps}
\end{minipage}
\hfill
\caption{Comparison of numerically calculated
nonlinear order parameter profiles $u$ and
$v$ for $\{110\}$ crystal faces obtained from the PFC
amplitude equations (dashed line) and the GL theory \cite{Wuetal06}
(solid line) and computed form MD simulations
with $\vec K_{101}$ and $\vec K_{110}$ for
$u$ and $v$, respectively (solid circles).
}
\label{uandvfull}
\end{figure}

The predictions of the different continuum theories are compared to
MD simulations in Table II. Interestingly, despite the differences in
quartic coefficients described in section \ref{subsec:quartic},
the PFC amplitude equations and GL theory give essentially identical
predictions.  The density wave profiles predicted by the two theories are almost 
indistinguishable on the scale of Fig. \ref{uandvfull}. Furthermore, the
predicted $\gamma$ values by the different continuum theories 
for a given crystal face do not differ by more 
than a few tenth of a percent. Both 
theories predict a weak four-fold anisotropy
$\epsilon_4\equiv (\gamma_{100}-\gamma_{110})/(\gamma_{100}+\gamma_{110})$ 
close to one percent consistent with the
results of MD simulations with the MH(SA)$^2$ EAM
potential \cite{MHSA2} for Fe. 

The PFC simulations predict essentially
the same anisotropy value but about 10\% larger $\gamma$ values that are in
closer agreement with MD simulation results. The larger $\gamma$ values
can be attributed to larger $|\vec K|$ modes and to the variation of 
the mean density in the interface region, both of which are neglected in
the weakly-nonlinear amplitude equations and GL theory. 

\begin{table*}[h]
\caption{Comparison of $\gamma$ values for different
crystal faces (in erg/cm$^2$) and anisotropy parameters including
$\epsilon_4\equiv (\gamma_{100}-\gamma_{110})/(\gamma_{100}+\gamma_{110})$
in percent
and $\epsilon_1$ and $\epsilon_2$ values (see text), predicted by MD simulations, 
and by various continuum theories (PFC simulations, PFC amplitude equations, and
GL theory) with the input parameters of Table I from MD simulations.} 
\centering
\begin{tabular*}{0.75\textwidth}%
     {@{\extracolsep{\fill}}ccccccc} \hline
 & $100$ & $110$ & $111$ &$\epsilon_4 (\%)$ & $\epsilon_1$ & $\epsilon_2$\\ 
\hline
MD (MH(SA)$^2$) (Ref. \cite{sun04})& 177.0 (10.8) & 173.5 (10.6) & 173.4 (10.6) & 
1.0(0.6) & 0.033 & 0.0025 \\ \hline
PFC simulation & 160.47 & 156.83 & 152.00 & 1.15 & 0.075 & -0.0094 \\ \hline
Amplitude equations & 144.14 & 140.67 & 135.76 & 1.22 & 0.082 & -0.0110 \\ \hline
GL theory \cite{Wuetal06} & 144.26 & 141.35  & 137.57& 1.02 & 0.066 & -0.0082 \\ 
\hline
\end{tabular*}
\label{tabgammas}
\end{table*}

The anisotropy parameter $\epsilon_4$ defined in terms of $\gamma_{100}$ and 
$\gamma_{110}$ has been traditionally used to quantify the magnitude of
anisotropy in dendrite growth theory  
\cite{Lan87,Kesetal88,BenBre93,KarRap9698,Provetal98}. 
As seen in Table II,
this parameter is reasonably well predicted by the PFC simulations and
amplitude equations or GL theory. Over the past few years, however,
numerous MD simulation studies have consistently found that at least
two anisotropy parameters $\epsilon_1$ and $\epsilon_2$ are necessary
to represent the entire $\gamma$-plot of fcc-liquid 
and bcc-liquid interfaces in diverse systems \cite{Hoyetal03}. 
These parameters are 
defined by the expansion of $\gamma$ in terms of 
cubic harmonics (i.e., combination of spherical
harmonics with cubic symmetry) that has the form
\begin{eqnarray}
\gamma(\hat 
n)&=&\gamma_0\left[1+\epsilon_1\left(\sum_{i=1}^3n_i^4-\frac{3}{5}\right)
\right.\nonumber\\
& 
&+\epsilon_2\left.\left(3\sum_{i=1}^3n_i^4+66n_1^2n_2^2n_3^2-\frac{17}{7}\right)
\right],
\end{eqnarray}
where the $n_i$'s are the coordinates of the direction 
normal to the interface ($\hat n$) in a set of cartesian
coordinates parallel to the crystal axes. Values of $\gamma$
for the the three independent crystal faces listed
in Table II uniquely fix $\gamma_0$,
$\epsilon_1$, and $\epsilon_2$. 
While a positive  
$\epsilon_1$ favors dendrite growth along the set of
six $\langle 100 \rangle$ directions,
a negative $\epsilon_2$ favors growth along the set
of twelve $\langle 110 \rangle$ directions. 
A recent phase-field simulation study has revealed the existence of
hyper-branched dendrite morphologies with a basic
set of twenty four growth directions between $\langle 100 \rangle$
and $\langle 110 \rangle$ over
some region of the $(\epsilon_1,\epsilon_2)$ parameter
space, where $\epsilon_1>0$ and $\epsilon_2<0$ favor
different growth directions \cite{Haxetal06}.

As seen from Table II, the agreement between
the different continuum theories and MD simulations is poorer
for the ratio $\gamma_{111}/\gamma_{100}$ than for
$\gamma_{110}/\gamma_{100}$. Consequently, the
$\epsilon_1$ and $\epsilon_2$ values, which depend
on these two ratios, are not well predicted
by these theories in comparison to $\epsilon_4$, which depends only on
$\gamma_{100}/\gamma_{110}$. 
This discrepancy appears to be an intrinsic limitation 
of weakly nonlinear theories where 
anisotropy is computed using only one set of density waves 
associated with principal reciprocal lattice vectors of
magnitude $K_{max}$.
While this one-set approximation is reasonably good
on the liquid side of the interface, where the density
wave amplitude is small, it breaks down on the solid side
where the highly nonlinear crystal density 
field is better approximated by sharply peaked Gaussians 
centered around atomic positions.  
Resolving this field requires a very large 
number of sets of reciprocal lattice vectors 
\cite{Laietal87}. 

An interesting related issue is the sensitivity
of crystalline anisotropy to microscopic details
of interatomic potentials. MD simulations to date indicate that
the magnitude of this anisotropy tends to be larger for fcc 
than bcc forming systems, suggesting that crystal structure is a main
determinant of anisotropy. Despite this trend, anisotropy values do
depend on the choice of potentials for a given crystal
structure. For example, two other interatomic potentials for Fe 
yield values of $\epsilon_4$ twice smaller than
for the MH(SA)$^2$ potential \cite{Hoyetal03}.

In contrast, anisotropy values are independent of
material parameters in both the PFC amplitude equations and
GL theory. The reason is that all the material-dependent
input parameters, which include the
density wave amplitude in the solid $u_s$ and
peak liquid structure factor properties,  
can be scaled out of the free-energy functionals for
these theories. This is readily seen in the dimensionless
form of the free-energy functional for the PFC
amplitude equations given by Eq.~(\ref{eq:amp_pot}).
Consequently, the ratios of $\gamma$ values for different
crystal faces that determine the anisotropy parameters
$\epsilon_1$ and $\epsilon_2$ are universal for 
all bcc elements within the confines of each theory,
and the value of anisotropy parameters depend on the  
nonlinear coupling between density waves.  
The results of Table II show that differences in 
these couplings (i.e., coefficients of quartic terms 
in the free-energy functionals) lead to only small
differences of anisotropy values. It is possible, however,
that other choices could produce values of $\epsilon_1$ 
and $\epsilon_2$ in closer agreement with MD simulations.

\section{Conclusions}

We have studied equilibrium properties of  
bcc-liquid interfaces in a physically motivated
small $\epsilon$ limit of the PFC model \cite{Eldetal02,Eldetal04}
where the freezing transition is weakly first-order.
This limit lends itself naturally to 
a multiscale analysis that was used to derive a set
of equations for the leading order
amplitudes $A_i^0$ of density waves corresponding to
the set $\{\vec K_i\}$ of
principal reciprocal lattice vectors, and to
express the free-energy of the solid-liquid
system as a functional of these amplitudes. Furthermore,
by exploiting the close analogy between this functional and
GL theory derived from classical DFT \cite{Shietal87,Wuetal06},
we have determined all the parameters of the PFC model
in terms of peak properties of the liquid structure factor 
and the solid density wave amplitude. 

In both the PFC amplitude
equations and GL theory, the anisotropy of $\gamma$ originates from the
directional dependence (i.e., the dependence on $\vec K_i\cdot \hat n$
where $\hat n$ in the interface normal) of the coefficients of
gradient-square terms ($|\vec \nabla A_i^0|^2$)
in the free-energy functional, which
govern the spatial decay rate of density waves in the liquid. 
In the isotropic limit where this directional dependence is neglected,
and hence all amplitudes are equal, $A_i^0=\phi$ for all $i$, 
both theories reduce to the conventional phase-field model
of solidification formulated in terms of the non-conserved
order parameter $\phi$. From this standpoint, the present analysis 
relates formally the crystal and conventional phase-field models. 

Numerical results show that the PFC model, the amplitude equations derived
from this model, and GL theory \cite{Wuetal06} all give very similar 
predictions of $\gamma$ and its anisotropy for parameters
of Fe where $\epsilon$ is small enough ($\epsilon\approx 0.1$) for the
amplitude equations to be quantitatively valid. The magnitude
of $\gamma$, the shape of density wave profiles in the spatially
diffuse solid-liquid interface region, and the standard crystalline
anisotropy parameter $\epsilon_4$ defined in terms of the ratio 
$\gamma_{110}/\gamma_{100}$, are in good overall agreement
with the results of MD simulations. The various continuum theories, however,
do not predict accurately higher order anisotropies that also depend on the ratio 
$\gamma_{110}/\gamma_{100}$. These anisotropies probably depend generally on  
the contributions of higher sets of reciprocal 
lattice vectors, which are neglected in the simplest 
formulation of the PFC model considered here.

\begin{acknowledgments}
This research was supported by U.S. 
DOE through Grants No. DE-FG02-92ER45471
as well as the DOE Computational 
Materials Science Network program. We thank Jeff Hoyt
and Mark Asta for valuable exchanges.
One of us (K.W.) wishes to thank M. P. Gururajan for helpful discussions.
\end{acknowledgments}

\end{document}